\documentclass[runningheads]{svmult}

\usepackage{makeidx}   
\usepackage{graphicx}  
\usepackage{subeqnar}  
\usepackage{multicol}  
\usepackage{physprbb}  
\makeindex             

\begin{document}
\title*{Brightest cluster galaxy formation in the cluster
C0037-2522: flattening of the dark matter cusp}
\toctitle{Brightest cluster galaxy formation in the cluster
C0037-2522: flattening of the dark matter cusp}

%
%
\titlerunning{Brightest cluster galaxy formation and flattening of the
dark matter cusp}
%
\author{Carlo Nipoti\inst{1,2}
\and Massimo Stiavelli\inst{3}
\and Luca Ciotti\inst{2}
\and Tommaso Treu\inst{4}
\and Piero Rosati\inst{5}}
\authorrunning{Carlo Nipoti et al.}
%
%
\institute{Theoretical Physics, 1 Keble Road, Oxford, OX1 3NP, UK \and
Dept. of Astronomy, University of Bologna, via Ranzani 1, 40127 Bologna, Italy \and
Space Telescope Science Institute, 
      3700 San Martin Drive, Baltimore, MD 21218 \and
Dept. of Physics \& Astronomy, UCLA, Box 951547, Los Angeles, CA 90095-1547 
\and
ESO, Karl-Schwarzschild-Strasse 2, 85748 Garching, Germany
}

\maketitle              

\begin{abstract}
The X-ray cluster C0337-2522 at redshift $z=0.59$ hosts in its core a
group of five elliptical galaxies. Using N-body simulations we show
that a multiple merging event among the five galaxies is expected to
take place in the next few Gyrs, forming a central brightest cluster
galaxy. We also find indications that dynamical friction heating
associated with this event is likely to modify the central slope of
the cluster dark matter density profile.
\end{abstract}

\section{Introduction}

We have identified a group of five ellipticals (Es) located in the
core of the X-ray cluster C0337-2522 at redshift $z=0.59$ (ROSAT Deep
Cluster Survey~\cite{rosati}). This system represents a strong
candidate for the initial stages of brightest cluster galaxy (BCG)
formation through galactic cannibalism~\cite{ostriker}.  Here, we
explore how many of the five galaxies under consideration are expected
to merge before $z=0$ and we study the properties of the merger
remnant.  We address these questions by using N-body simulations,
exploring several initial conditions compatible with the imaging and
kinematic information from our ESO-VLT data. The details of
observations and numerical simulations are given in~\cite{nipoti}.

Recent observational studies~\cite{sand1,sand2} of a few galaxy
clusters find central dark matter (DM) density distributions
$(\rho_{\rm DM}~\propto~r^{-\beta})$ flatter ($\beta \sim 0.5$) than
predicted by cold dark matter simulations~\cite{navarro,moore} ($\beta
\sim 1-1.5$). A possible interpretation of this discrepancy is that
dynamical friction heating~\cite{bertin} is effective in flattening
the DM cusp. We investigate this hypothesis using our simulations,
where the initial DM profile is cuspy ($\beta=1$), galaxies are
deformable, and initial conditions correspond to an observed cluster.

\begin{figure}[t]
\begin{center}
\includegraphics[width=.4\textwidth]{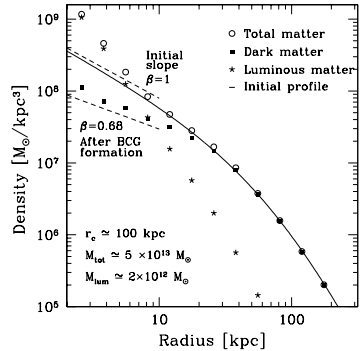}
\end{center}
\caption[]{Final DM, stellar, and total density profiles for a
representative simulation.  In this case, the best--fitting inner
slope of the final DM profile is $\beta=0.68$. The solid curve is the
initial ($\beta=1$) cluster DM profile.}
\label{eps1}
\end{figure}

\section{Results}
 
In all the simulations 3 to 5 galaxies merge before $z=0$. The merger
remnant is similar in its main structural and dynamical properties to
a real BCG (but with no evidence of the diffuse luminous halo typical
of cD galaxies). In particular, it satisfies the
Faber-Jackson~\cite{faber} relation and the K-band Fundamental
Plane~\cite{pahre}, under the hypothesis that the five Es are placed
on these scaling relations, and that the mass--to--light ratio remains
unchanged in the merging process.  However other features of Es, such
as the $M_{\rm BH}$-$\sigma_{0}$ relation~\cite{gebhardt,ferrarese},
and the metallicity gradient~\cite{peletier}, are hardly reproduced by
this multiple dissipationless merging scenario.  As regards the
properties of the cluster DM halo, we find final profiles flatter than
the initial $\rho_{\rm DM} \propto r^{-1}$ profile (see
Fig.~\ref{eps1}).  We fit the cluster DM density profiles with the
formula $\rho_{\rm DM} = \rho_{\rm DM,0}(r/r_{\rm c})^{-\beta}
[1+(r/r_{\rm c})]^{-4+\beta}$, with $\beta$, $r_{\rm c}$, and
$\rho_{\rm DM,0}$ free parameters.  For the final DM profiles in our
simulations $\langle~\beta~\rangle~\simeq~0.66~\pm~0.15$ to be
compared with $\beta=1$ of the initial profile. Our results (in
accordance with recent simulations using rigid galaxy
models~\cite{elzant}) point towards a major role of dynamical friction
associated with BCG formation in determining the cluster DM profile on
small scales.

%

\end{document}